\documentclass[utf8]{frontiersHLTH} 

\usepackage{url,hyperref,lineno,microtype,subcaption}
\usepackage[onehalfspacing]{setspace}

\def\keyFont{\fontsize{8}{11}\helveticabold }
\def\firstAuthorLast{Blair {et~al.}} 
\def\Authors{Johnna Blair\,$^{1,*}$, Jeff Brozena\,$^{1}$, Mark Matthews\,$^{2}$, Thomas Richardson\,$^{3}$ and Saeed Abdullah\,$^{1}$}
\def\Address{$^{1}$Penn State University, PA, USA \\
$^{2}$University College Dublin, Ireland\\
$^{3}$University of Southampton, UK}
\def\corrAuthor{Johnna Blair}

\def\corrEmail{jlb883@psu.edu}

\begin{document}
\onecolumn
\firstpage{1}

\title[FinTech for mental health]{Financial technologies (FinTech) for mental health: The potential of objective financial data to better understand the relationships between financial behavior and mental health} 

\author[\firstAuthorLast ]{\Authors} 
\address{} 
\correspondance{} 

\extraAuth{}

\maketitle

\begin{abstract}
Financial stability is a key challenge for individuals with mental illnesses. Symptomatic periods often manifest in poor financial decision-making including compulsive spending and risky behaviors. This article explores research opportunities and challenges in developing financial technologies (FinTech) to support individuals with mental health. Specifically, we focus on how objective financial data might lead to novel mental health assessment and intervention methods. We have used data from one individual with bipolar disorder (i.e., an N=1 case study) to illustrate feasibility of collecting and analyzing objective financial data alongside mental health factors. While we have not found statistically significant trends nor our findings are generalizable beyond this case, our approach provides an insight into the potential of using objective financial data to identify early warning signs and thereby, enable preemptive care for individuals with serious mental illnesses. We have also identified challenges of accessing objective financial data. The paper outlines what data is currently available, what can be done with it, and what factors to consider when working with financial data. We have also explored future directions for developing interventions to support financial wellbeing and stability. Furthermore, we have described the technical, ethical, and equity challenges for financial data-driven assessments and intervention methods, as well as provided a broad research agenda to address these challenges.



\tiny
 \keyFont{ \section{Keywords:} Mental Health, Financial technologies, FinTech, Open banking, Intervention, Impulsive spending, Privacy preserving} 
\end{abstract}

\section{Introduction}

Mental illness is a serious public health crisis on a global scale. It affects more than one billion individuals \cite{trautmann2016economic}. This results in significant economic consequences, with an estimated total annual cost of \$2.5 trillion globally \cite{trautmann2016economic}. This growing issue impacts many facets of daily life, including a strong association between mental health and financial instability. Individuals with mental health issues are more likely to live in relative poverty \cite{richardson2013relationship, saraceno1997poverty}. Symptoms of some illnesses can also manifest in poor financial decision making. More specifically, bipolar disorder (BD) appears to be linked with greater risk of impulsive financial behaviors---The diagnostic criteria for hypomanic and manic episodes specifically lists impulsive spending as possible symptoms \cite{american2013dsm}. Those with BD are at greater risk of problem gambling \cite{jones2015gambling}, and 71\% report impulsive spending whilst hypomanic \cite{fletcher2013high}. This impulsive spending has also been shown to be linked to negative feelings and subsequent comfort spending \cite{richardson2017relationship,richardson2018financial}. As this illustrates, there is often a cyclical and bidirectional relationship between financial stability and mental health---financial instability can worsen mental health, which in turn, can cause further financial challenges, leading to a vicious cycle \cite{richardson2015longitudinal, richardson2017longitudinal, richardson2017relationship, richardson2018financial}.

The lack of access to objective financial data has been a key challenge in understanding the nuanced relationship between mental health and financial behaviors. Prior studies have mostly used surveys \cite{richardson2013relationship} and focus group interviews \cite{richardson2017longitudinal} to assess financial behaviors associated with serious mental illnesses. This means that much of what we know about this relationship is mediated through subjective methods that may be prone to bias and retrospective recall error. For example, Richardson et al. \cite{richardson2018financial} measured self-reported compulsive spending in Bipolar Disorder, rather than objective data on spending patterns. While this provides a useful broad overview, there remains a knowledge gap regarding how idiosyncratic, context-driven, and illness-specific factors impact financial decision making and stability of an individual living with mental illness. Furthermore, the lack of granular, in-situ assessment methods is a key barrier against developing just-in-time, adaptive, and personalized interventions focusing on financial stability for this population. Given the importance of financial wellbeing for mental health, this remains a serious knowledge gap with broad practical implications.

In recent years, there has been considerable progress toward more open and accessible financial data. Financial institutes are increasingly adopting open banking application programming interface (API) mandating access to accounts, payments, and transactions. There have also been third-party tools and platforms (e.g., Plaid \cite{plaid}) that enable access to financial data across a broad range of institutes. We argue that the granular and (near) real-time access to individual financial data can lead to a paradigm shift in the domain of financial wellbeing and mental health. That is, it can provide a unique opportunity to explore the nuanced relationship between money and mental health, uncover new financial patterns indicative of early-warning signs, and develop preemptive interventions to support healthier financial decisions.

However, there are considerable challenges before we can achieve this vision of financial technologies (FinTech) to effectively support mental health. Specifically, there is a need to identify potential features and analytical methods that can leverage financial behavioral data to identify early-warning signs and opportune moments for intervention delivery. It will also be essential to design effective interventions that balance the need for personal agency and long-term financial stability. Addressing privacy concerns will be critical given the sensitive nature of the data --- a recent survey of those with bipolar disorder found that 97.5\% used smartphone apps but concerns about privacy were common \cite{morton2021using}. Individuals with mental illnesses might need support from others (e.g., family members) to make sustainable financial decisions. As such, the assessment and intervention methods should focus on shared, collaborative financial decision making.

Toward this broad research vision, we make the following contributions for future FinTech development:

\begin{itemize}
\item Discuss the potential for technology to serve as a useful way to map the link between finances and mental health and identify personalized warning signs indicating mental health issues.
\item Use a N=1 case study with financial data from an individual with bipolar disorder to illustrate the opportunities and technical challenges for collecting and using objective banking data in mental health contexts, such as data access and data quality.
\item Lay the landscape for future research on the potential intervention strategies and ethical challenges for using objective data in the financial mental health domain.
\end{itemize}

\section{Related Work}
\subsection{Financial Behaviors Related to Mental Health}
Mental health issues can lead to problematic financial behaviors. For instance, those with bipolar disorder have reported often avoiding finances and finding it hard to plan especially when depressed \cite{richardson2017relationship}. Depression has also been linked with cognitive impairments such as in attention, working memory and problem solving\cite{marazziti2010cognitive}, all of which have the potential to impact financial capability. In bipolar disorder, problematic financial behaviors including impulsive spending have been linked to impulsivity, low self-esteem, and thoughts around achievement and dependency on others \cite{cheema2015assessing, richardson2019financial}.This combination of purchasing behaviors related to manic and depressive mood episodes can have a significant impact on long-term personal finances. In the following section, we discuss prior work establishing problematic financial behaviors in relation to mental health conditions like bipolar disorder. These spending behaviors are often interrelated and can take on a cyclical pattern\cite{richardson2019financial}.

\subsubsection{Non-Purchasing Behaviors}
Many mental health conditions come along with decision-making challenges, memory difficulties, and avoidance behaviors that can have an impact on finances \cite{alexander2017reported, marazziti2010cognitive}. Memory or attention challenges which are common with some mental illnesses (e.g., depression, bipolar disorder) can make it difficult to remember important deadlines or due dates \cite{marazziti2010cognitive}. Without significant support or reminder systems, individuals may miss these deadlines and face overdue bills or late fees that negatively impact their finances. Different avoidance behaviors can affect financial decisions as well. Some may find it difficult to make a decision or feel anxious when presented with too many options \cite{park2018not}, so they may choose to put off making imperative financial decisions. Poor perceived financial wellness can increase stress and anxiety about finances, which can then lead to avoidance of financial matters and ultimately making their financial situation worse \cite{richardson2019financial}. Some may avoid acting on their finances entirely, like in cases of debt-related stress, which can also lead to overdue bills. If this avoidance results in neglecting to review banking balances, this can lead to negative outcomes, such as overdrawn accounts\cite{narayan2011role}. All these behaviors, though not focused directly on spending money, can make managing finances and budgeting even more difficult. 

\subsubsection{Compulsive Spending}
One of the leading issues in the context of mental health and money is compulsive spending, which is commonly reported in bipolar disorder \cite{richardson2017longitudinal} This can be a standalone condition or a symptom of other conditions like BD, creating further challenges to mental wellbeing. Compulsive buying, which can often present during manic mood episodes of BD, is an intense, irresistible need to spend beyond what is necessary, and in many cases beyond one’s financial means \cite{granero2016compulsive}. This can occur despite the emotional distress it may cause, limited funds, or a lack of actual need for the items purchased. Like other addictive behaviors, the amount of spending or the size of purchases may escalate over time as one needs more to feel the same level of outcome. For many, it is the desired part of a dopamine loop and a distraction from negative feelings \cite{granero2016compulsive}. While these purchasing habits may have temporary and fleeting positive outcomes for the individual, they can create increased feelings of regret, guilt, and even suicidality in the long-term \cite{richardson2017relationship}. Not only can they affect personal finances, but can also become a stressor in relationships and other parts of their life. This shame and disappointment associated with spending may then be followed by additional spending \cite{richardson2019financial}. 

\subsubsection{Symptom-Specific Spending}
Other BD symptoms can manifest in the types of purchases individuals choose to make. These include impulsivity, risk-taking behavior, goal-directed activity, and comfort spending. Whereas compulsive spending is a drive to spend money for various reasons, \emph{impulsive spending} is a momentary decision to make a purchase \cite{ramirez2020impulsivity}. Manic mood episodes in BD can result in high levels of impulsivity, as well as difficulty delaying gratification \cite{american2013dsm, ramirez2020impulsivity}. This can lead individuals to make quick decisions with money without considering the consequences of those decisions \cite{cheema2015assessing}. Increased \emph{risk-taking behavior} is another common costly symptom that manifests in financial decisions \cite{cook2004assessing, fletcher2013high, ramirez2020impulsivity}. These financial decisions could be perceived as high-risk, high-reward opportunities (e.g., investments or risky business ventures) that individuals jump into quickly, which can negatively impact finances in the long term \cite{cook2004assessing} and increase regret and feeling of guilt \cite{fletcher2013high}. Some may also see an increase in \emph{goal-directed activities}, where they are very focused on completing tasks toward a specific project or multiple big ideas \cite{american2013dsm}. Those with bipolar disorder tend to have ambitious goals linked to manic symptoms \cite{tharp2016goals}, and cognitions around achievement predict more spending \cite{richardson2019financial}. This can be activities like starting a new business or learning a new hobby, which often include many upfront costs to support. 

On other occasions, depressed mood episodes may give way to \emph{comfort spending} \cite{richardson2017relationship}. These purchases are made as an attempt to change one’s mood state or to give themselves something to look forward to in the future \cite{richardson2017relationship}. This may also take the form of charity or gifts for others, rather than a purchase for themselves. In this case, this can be seen as improving their mood or reducing guilt about their purchases by being generous to others \cite{richardson2019financial}. Research with BD found that symptoms of depression, stress, and anxiety increased impulsive spending over time, linked with this theme of ‘comfort spending’ as an outcome\cite{richardson2017relationship, richardson2018financial}. A model based on bipolar disorder by Richardson et al.\cite{richardson2019financial} suggests an interpersonal vicious cycle where regret and guilt from impulsive or high-risk spending leads to worries about others and excessive generosity as a way to cope with this, thus fueling more spending. This suggests that these symptom-related behaviors can be connected. Additional research involving real-time purchasing data could help uncover other relationships and driving factors between different types of spending behaviors.

\subsubsection{Patterns of Spending}
Previous work has provided insight into the different goals and drivers behind spending in the context of BD, less is known about the specific patterns that may exist in these spending behaviors. For some, impulsive spending may result in a single high-dollar purchase \cite{fletcher2013high, ramirez2020impulsivity}. Others may experience spending sprees or bursts of many items purchased within a narrow window of time, as is common with bipolar disorder \cite{american2013dsm}. While these temporal bursts of spending have been explored in regards to individual psychological traits (e.g., neuroticism \cite{tovanich2021eds}), however spending patterns in relation to symptoms of mental illness have been under-explored. Our current understanding of how spending temporally presents in relation to symptoms still lacks the depth needed to help predict these behaviors and develop personalized interventions to provide preemptive support.

While previous studies have resulted in valuable insights on how symptoms can manifest in specific financial behaviors, they might not lead to the complete picture given their reliance on self-reported perceptions of individual’s own behaviors. In other words, self-reported data can suffer from incomplete recall and other biases, particularly during symptomatic periods. Being able to access objective personal financial data can help to address this issue. That is, we can use granular and objective financial data spanning symptomatic and non-symptomatic periods to identify idiosyncratic patterns indicative of problematic financial behaviors. However, there are considerable challenges in accessing objective financial data. In the following section, we discuss strategies and highlight these challenges using a N=1 case study in accessing financial data from an individual with bipolar disorder.

\section{Financial Behaviors and Bipolar Disorder: N=1 Case Study}
As previous research illustrates, the relationship between mental health and money is highly complex and nuanced. In this paper, we focus on bipolar disorder and spending behaviors as an example case to investigate this further. Symptoms related to BD impact several facets of daily life, including personal finances. BD is typically characterized by cycling between the highs and lows of manic and depressive mood episodes. A depressive mood episode often includes severe feelings of sadness, low-energy, indecisiveness, and poor concentration \cite{american2013dsm}. Conversely, manic mood episodes often feature feelings of euphoria, increased energy, high activity, risk-taking behaviors, impulsivity, and a decreased need for sleep \cite{american2013dsm}.

These different mood episodes provide a framework to better understand the relationship between mood and money and the spending patterns that present during different mental health contexts more broadly. This can also inform the ideal points for intervention to help support healthy financial choices. However, there are considerable challenges in accessing objective financial data. In the following section, we discuss strategies and highlight these challenges using a N=1 case study in accessing and processing financial data from an individual with bipolar disorder. 

\subsection{Methods}
We conducted a case study involving a single individual, also a coauthor, who experienced a relapse into a hypomanic state lasting between 2017 and 2018. This episode was the individual's first substantial symptomatic period since their 2008 diagnosis of bipolar disorder type II. In the following section, we discuss the process of collecting, pre-processing, and analyzing granular financial data across symptomatic and non-symptomatic periods.

\subsubsection{Label Creation}
To determine symptomatic periods during the study duration, we used the National Institute of Mental Health’s Life-Chart Method (NIMH-LCM) \cite{leverich2002}. Prior work has validated the use of NIMH-LCM for longitudinal assessment of bipolar disorder \cite{honig2001peac}. Our coauthor collaborated with their family members to complete a Retrospective NIMH-LCM Self-Rating form for a period spanning 2017 to 2018. Periods of hypomania were logged as mild, moderate, or severe following the form's criteria.  Depressive symptoms were also measured on a similar scale.

During the study period, the co-author and their family members recalled that the hypomanic symptoms had escalated slowly with no sudden variations in mood over time. They chose to represent these slower variations in mood by marking their responses to the Retrospective NIMH-LCM on a monthly basis. They began by accounting for a small number of landmark events serving as anchor points in relation to any additional events they could recall. They supplemented their dialogue with emails, photographs, and SMS logs to produce additional date ranges where symptoms were present. These steps resulted in 2 periods of mild mania lasting 5 months in total, and 5 periods of moderate mania lasting 9 months in total. They did not identify any period with depressive symptoms during the study duration.

\subsubsection{Extracting Data from Financial Statements}
The banking institution associated with this case study does not provide a mechanism for access to structured financial data. As such, we had to manually recreate a structured financial dataset by programmatically extracting tabular text content of digitized financial statements (i.e., PDF files). We collected a total of 24 monthly statements as multipage PDF files spanning the study duration. Each PDF file contained transaction-level detail for all open accounts.

We first extracted the tabular text content of each statement into comma-separated text files using Camelot \cite{camelot}, an open-source Python package for PDF text extraction. We manually inspected output from Camelot to ensure accuracy. Specifically, we compared a series of 25 transaction dates, descriptions, and amounts against the original PDF statement, all of which were extracted correctly by Camelot. The formatting of the original PDF files was read with a similarly high degree of accuracy. While the text output was technically accurate, formatting issues prevented its immediate use. Camelot preserved all pagination and line breaks of the original statement, resulting in cases where a single transaction could span multiple rows, a single account could span multiple files, and multiple accounts could be contained within a single file.

\begin{figure}[h]
\begin{center}
\includegraphics[width=0.45\textwidth]{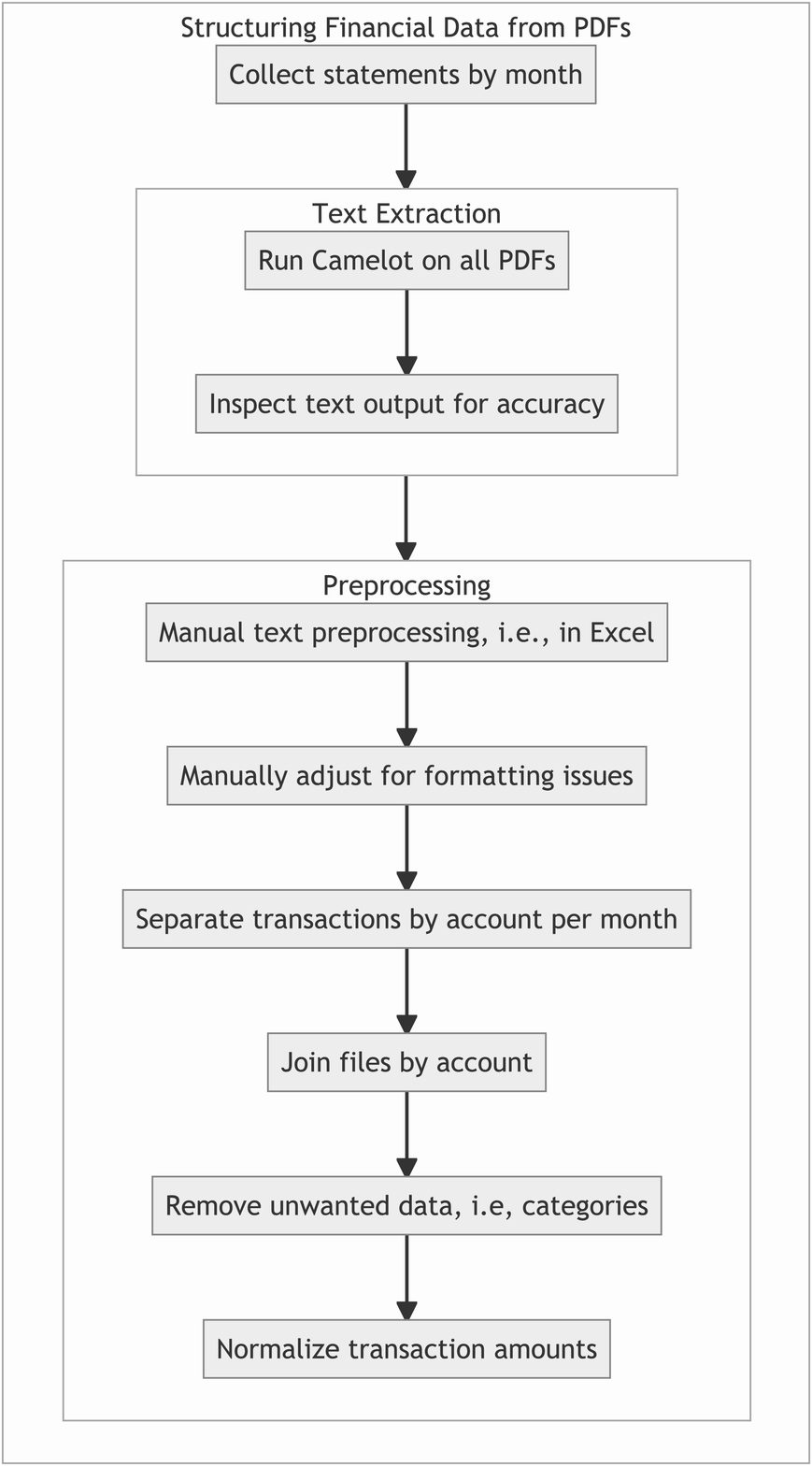}
\end{center}
\caption{\label{fig:flowchart}Workflow to create structured financial data from PDF statements}
\end{figure}

Accounting for the variety of these exceptions required a non-trivial amount of manual preprocessing in order to produce a dataset where each row contained a single, complete transaction. We used Microsoft Excel to visually inspect and clean all Camelot output. We then manually merged transactions spanning multiple rows into a single row. During the data processing, we retained account labels (i.e., checking and credit). We used these labels to separate transactions in a single file per account per month. We then  joined these files in sequence to produce a single file per account containing transaction dates, descriptions, and transaction amounts. We also sort the transactions by date. Figure \ref{fig:flowchart} summarizes the pre-processing steps.

We then used pandas \cite{pandas} --- an open source Python package --- to transform information in comma-separated files into a structured dataset. To preserve privacy, we removed all text-based transaction descriptions. Additionally, we normalized the transaction values between 0 and 1 per account. We hypothesized that problematic financial behaviors will manifest in expenditure transactions. As such, we focused on expenditure and omitted income-based transactions from our analysis. This resulted in creating a single \emph{privacy-preserving} dataset containing normalized expenditure transactions with labels indicating originating account but containing no personally identifiable information nor actual amount.

We performed our analysis based on this combined set using two strategies. The first strategy focused on analyzing details on transaction volume for all accounts, labeled by account and grouped by date. The second strategy used detail on transaction counts to analyze by frequency of expenditure, again labeled by account and grouped by date.

\subsubsection{Analysis}
We conducted an exploratory visual analysis of these two datasets to identify trends in frequency and amount of expenditure. Transactions were resampled on a daily, weekly, and monthly basis. For each of these resampled periods, we visualized transaction frequency and volume, visually shading each figure to indicate symptomatic severity following the NIMH-LCM labels generated earlier. We define transaction frequency as the number of transactions occurring in a given time period, and transaction volume as the normalized, aggregated dollar-value of purchases made during a time period.

To account for unbalanced group sizes and the possibility of unequal variances, we used a one-way Welch analysis of variance to determine how expenditure varied between symptomatic phases. Each transaction was marked as  “mild”, “moderate”, or “none” to account for symptom severity using the NIMH-LCM labels. We also performed a one-way Welch ANOVA after merging “mild” and “moderate” periods together. For each one-way Welch ANOVA test, a Games-Howell post-hoc test was performed to produce confidence intervals for the differences between group means and test for statistical significance.

We expected to see occasional increases in transaction frequency, hypothesizing that these would correspond to periods of hypomania and mania. Following methods outlined in Tovanich et al. \cite{tovanich2021eds}, we explored potentially impulsive, short-term increases to spending frequency by calculating a burstiness parameter. 
\[ B = \frac{r - 1}{r + 1}, \]
where \(r = \sigma / \tau\). Here \(\tau\) is the average and \(\sigma\) is the standard deviation of interevent timing for all transactions. Here, a burstiness parameter of 1 indicates spikes, 0 indicates randomness, and -1 indicates stability.
We calculated burstiness for the entire dataset. We also calculated burstiness for each symptomatic phase after grouping transactions by symptomatic intensity. We tested for significant differences in burstiness between phases using a one-way Welch ANOVA test with a Games-Howell post-hoc test.

\subsection{Findings}
\subsubsection{Exploratory Visual Analysis}
\begin{figure}[ht]
\begin{center}
\includegraphics[width=\textwidth]{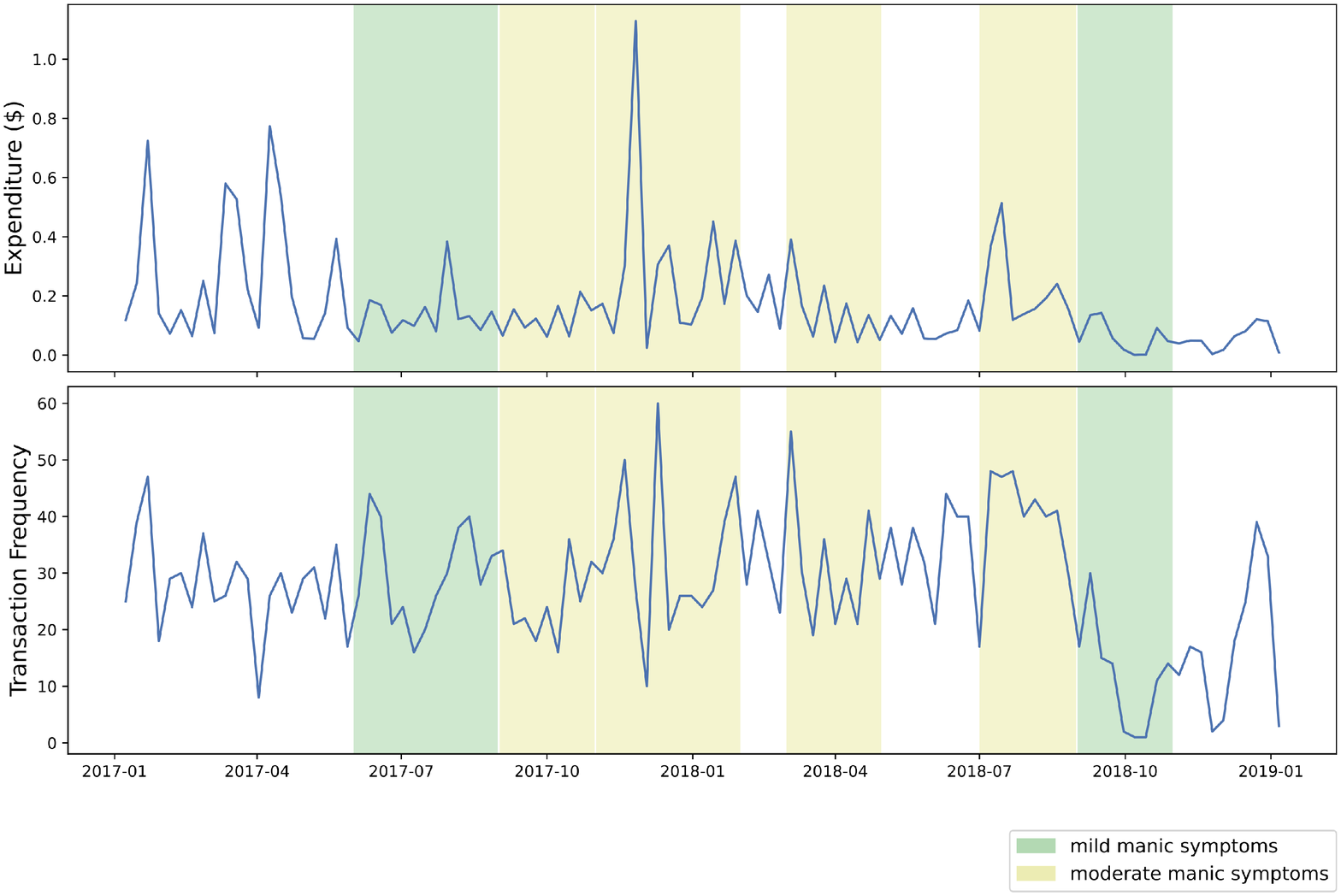}
\caption{\label{fig:weekly-checking}Weekly expenditure from transaction account.}
\end{center}
\end{figure}

\begin{figure}[ht]
\begin{center}
\includegraphics[width=\textwidth]{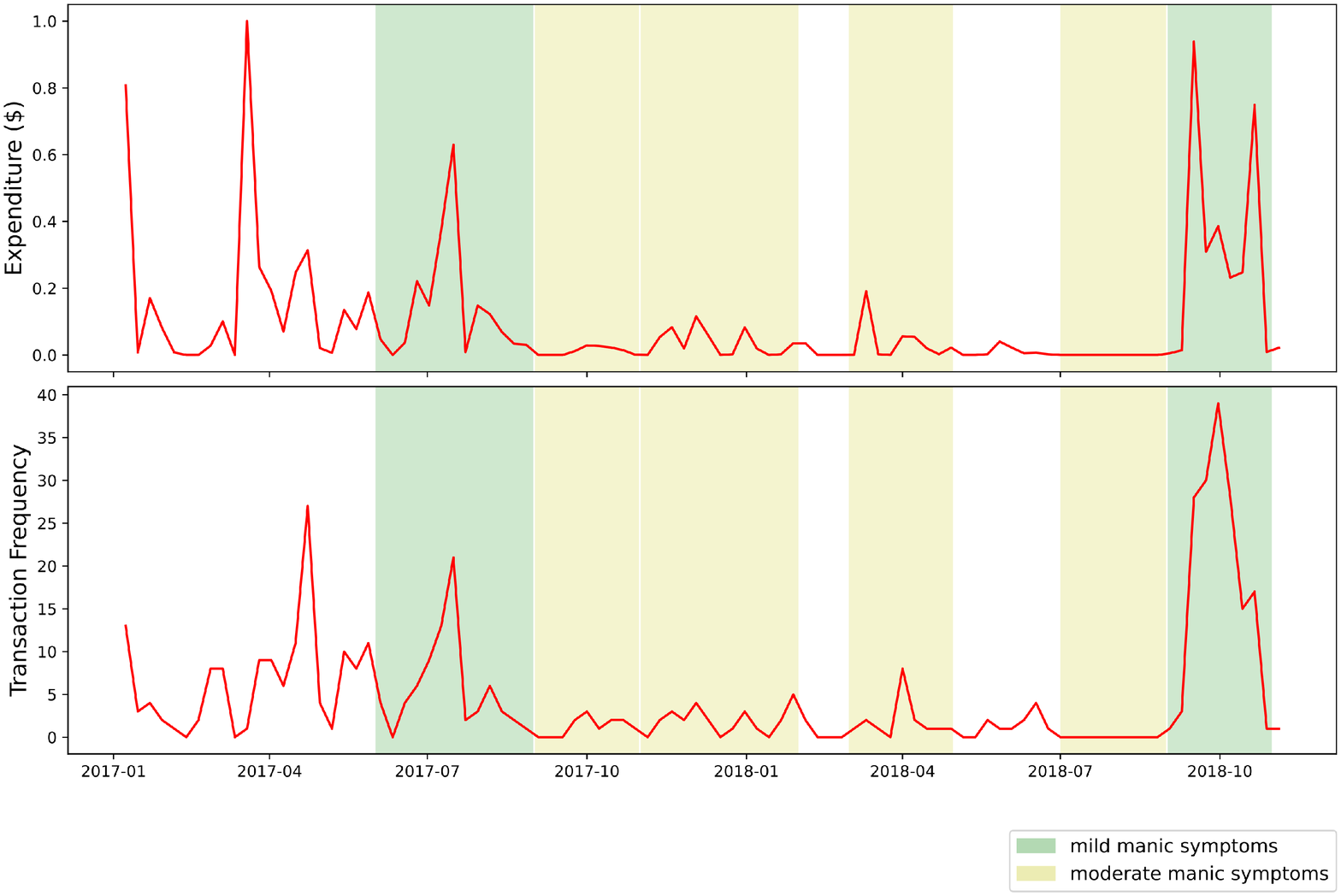}
\caption{\label{fig:weekly-credit}Weekly expenditure from credit account.}
\end{center}
\end{figure}

Our visual analysis of transaction volume and frequency during approximated periods of manic symptoms yielded several points of interest. In Figure \ref{fig:weekly-checking}, several noticeable spikes in expenditure from the primary transaction account are visible during periods of moderate manic symptoms, shaded in yellow. As Figure \ref{fig:weekly-credit} shows, two periods of mild manic symptoms, shaded in green, are marked by an increase to credit expenditure. A mixture of risky spending dynamics occurs during late 2018 where a marked decrease in cash-based spend coincides with an increase of credit spend in Figure \ref{fig:weekly-credit}).

\begin{figure}[ht]
\begin{center}
\includegraphics[width=\textwidth]{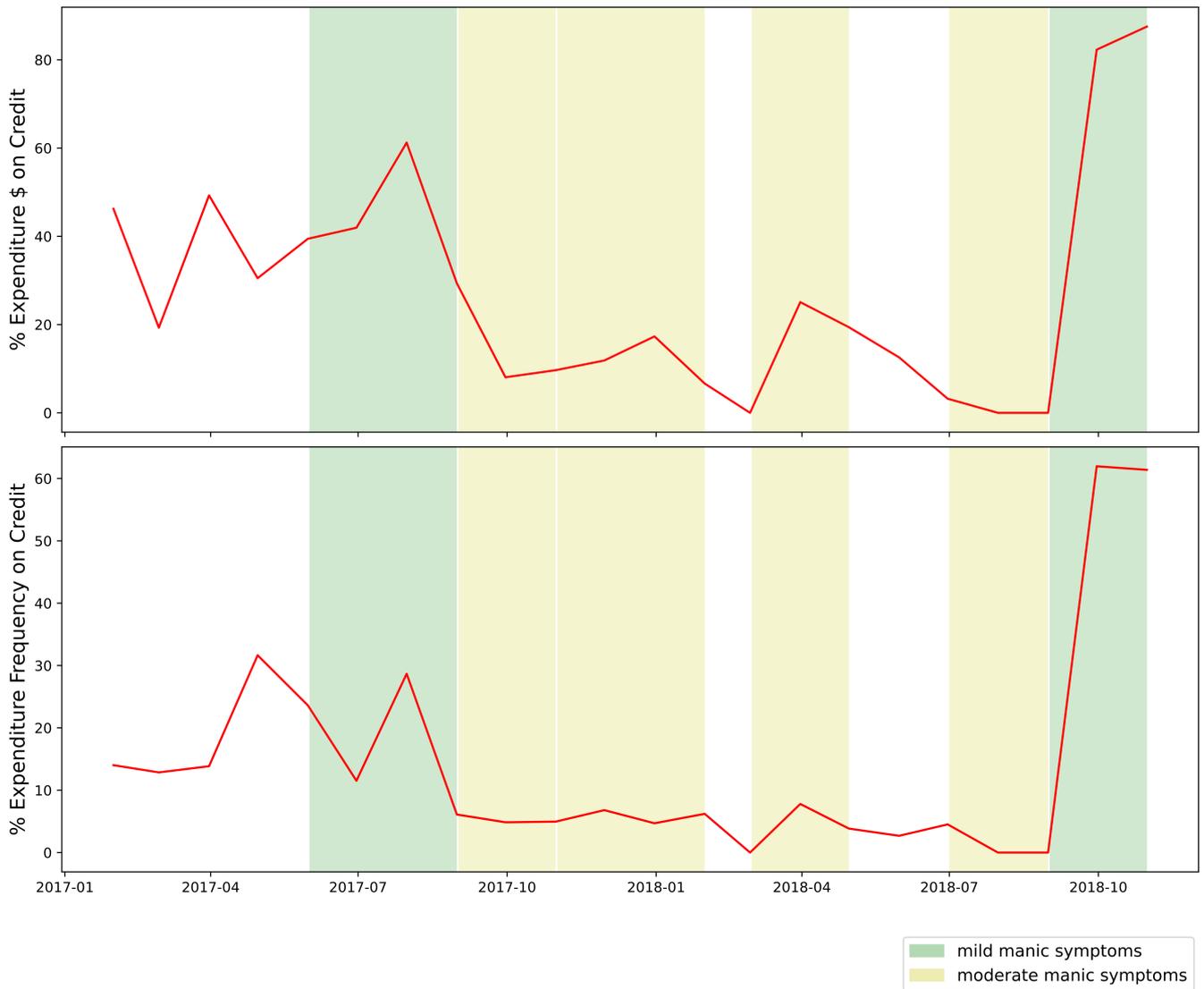}
\caption{\label{fig:monthly-credit}Monthly Percentage of Purchases Made on Credit}
\end{center}
\end{figure}

This case study intentionally excluded income-related transactions, assuming that many individuals have more control over expenditure than income. Although income-related data could provide interesting features (i.e., irregular income as a sign of partial employment or the stressors of gig work), we chose to focus our case study on the relationship between mood episodes and expenditure, including the use of credit. The complicated relationship between debt and bipolar disorder is illustrated in Figure \ref{fig:monthly-credit}, where the percentage of credit-based transactions is shown to increase as this symptomatic episode concludes.
\subsubsection{Quantitative Analysis}
There were a total of 3,373 transactions in our analysis. We grouped daily transaction data (frequency and volume) by intensity of symptoms (“mild”, “moderate”, “none”) and calculated the mean for each group. Our Welch ANOVA test yielded no statistically significant differences in transaction frequency ($F_{1,362} = 0.31, p = .734$) or volume ($F_{2,362} = 1.93, p = .147$). With transactions grouped by the presence or absence of any symptoms, we calculated the mean for each group. No statistically significant differences were noted in transaction frequency ($F_{1,558} = 0.35, p = .556$) or volume ($F_{1,431} = 0.19, p = .665$) between groups.

\begin{table}[]
\centering
\begin{tabular}{|l|l|l|l|}
\hline
 & None & Mild & Moderate \\ \hline
Frequency & 5.23 & 5.49 & 5.34 \\ \hline
Volume & 0.05 & 0.05 & 0.04 \\ \hline
\end{tabular}
\caption{Normalized mean daily amounts for transaction frequency and volume by symptom intensity.}
\label{tab:group-means}
\end{table}

\begin{table}[]
\centering
\begin{tabular}{|l|l|l|}
\hline
 & None & Sympomatic \\ \hline
Frequency & 5.23 & 5.39 \\ \hline
Volume & 0.05 & 0.04 \\ \hline
\end{tabular}
\caption{Normalized mean daily amounts for transaction frequency and volume by presence or absence of symptoms.}
\label{tab:group-means-2}
\end{table}

Following Tovanich et al. \cite{tovanich2021eds}, we calculated two different burstiness parameters: $B_{D}$ with the difference in days between each transaction as the interval, and $B_{C}$ with the number of consecutive days without any expenditure as the interval. In our dataset, we found $B_{D}$ to be 0.402 and $B_{C}$ to be -0.148. Note that a burstiness parameter of 1 indicates spikes, 0 indicates randomness, and -1 indicates stability. No statistically significant differences between symptomatic periods and non-symptomatic periods were found using either interval calculation.

These non-significant results may be due in part to the small sample size of this case study or to the sparse labels available to our dataset. This analysis could be improved with additional longitudinal financial data for comparison. If mood logs or specific health records were available, symptomatic periods could be labeled with greater accuracy. The presence of more detailed transaction data and metadata (i.e., timestamps, purchase locations) could significantly improve future analyses.
\begin{table}[]
\centering
\begin{tabular}{|l|c|c|}
\hline
 & \multicolumn{1}{l|}{None} & \multicolumn{1}{l|}{Symptomatic} \\ \hline
Mean $B_{D}$ & 0.40 & 0.38 \\ \hline
Mean $B_{C}$ & -0.43 & -0.53 \\ \hline
\end{tabular}
\caption{Mean burstiness parameters for periods with and without the presence of symptoms.}
\label{tab:burst-means}
\end{table}

\subsubsection{Unsupervised Anomaly Detection}
In addition to the above statistical testing, we implemented an unsupervised isolation forest algorithm \cite{liu2008} using scikit-learn \cite{scikit-learn} to detect potential anomalies in weekly spending frequency throughout the analysis period without relying on data labels. This isolation forest algorithm requires a user-defined ``contamination parameter" --- the estimated percentage of anomalous data points in a dataset. We chose a contamination parameter of $0.05$ following a visual inspection of values ranging from $0.01$ to $0.1$. Setting this parameter value below $0.05$ resulted in only the most extreme fluctuations in spending frequency being identified. Larger values resulted in the detection of an increasing number of false positives. Taking this approach to parameter selection was possible with existing background knowledge of the case study. 

\begin{figure}[ht]
\begin{center}
\includegraphics[width=\textwidth]{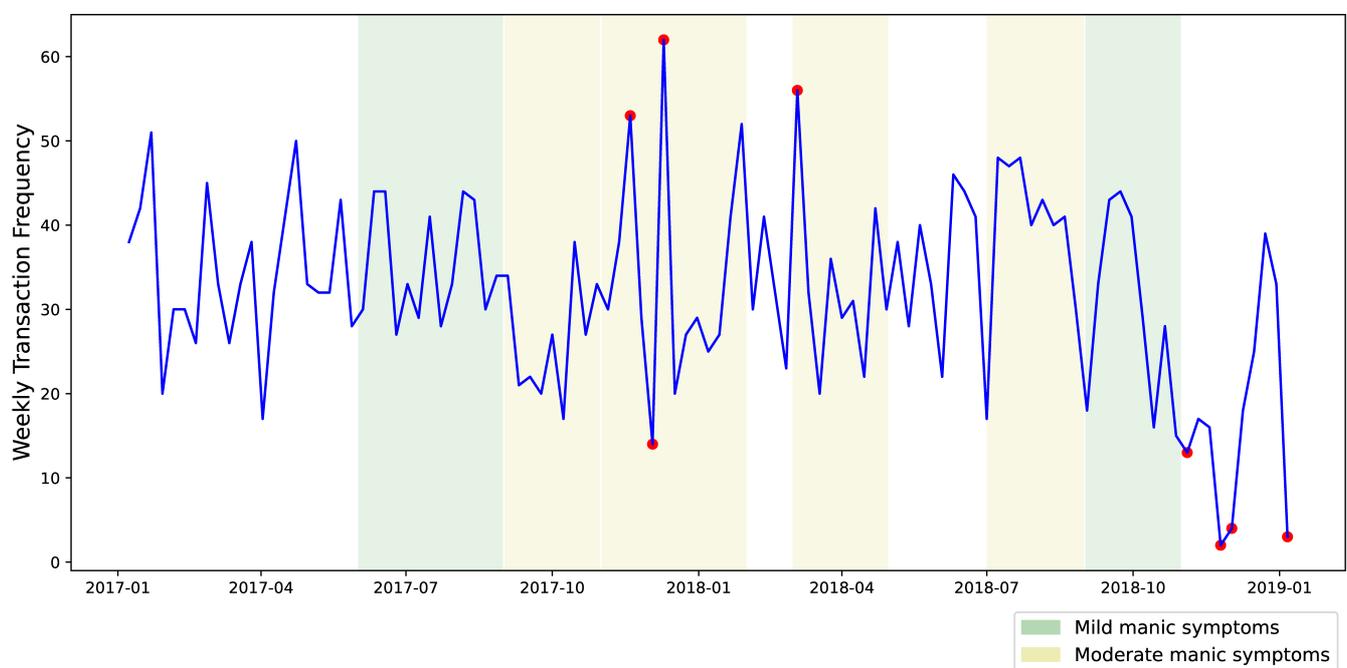}
\caption{\label{fig:weekly-anomaly}Outlier periods of spending frequency as identified by an unsupervised isolation forest algorithm.}
\end{center}
\end{figure}

Figure \ref{fig:weekly-anomaly} displays outlier periods of spending frequency detected by the isolation forest algorithm. In this case, both high and low frequency periods were identified by the algorithm. The majority of the higher frequency spending identified by the algorithm occurred during self-reported periods of moderate mania. Additionally, an anomalous decline in spending frequency was identified following the conclusion of this episode. Our approach shows the potential to identify relationships between financial behaviors and mental health states. A potential early warning system could integrate these results with other predictive approaches to form a more complete representation of risky financial behavior. 

\section{Discussion}
In this paper, our goal is to explore financial ``biomarkers” indicative of mental health issues as well as identifying opportunities and challenges in developing financial data-driven interventions. Toward this goal, we conducted a N=1 case study to collect, process, and analyze financial transactions from an individual with bipolar disorder to illustrate the potential, pragmatic considerations and factors to consider when working with objective financial data. From this dataset, we have identified potentially risky financial behavioral dynamics across symptomatic phases (e.g., ratio of increased credit spending). However, we have not found any statistically significant trends, which might be due to the small sample size (i.e., only 3,373 transactions spanning two instances of mild mania and five instances of moderate mania in total). Furthermore, our findings and analyses are limited to this N=1 case study, which might not generalize to individuals with bipolar disorder or other mental illnesses.

However, our approach illustrates the potential benefits and challenges of accessing, working with, and collecting ground truth associated with objective financial data. In the following sections, we will first discuss potentials for using objective financial data to detect early warning signs and interventions focusing on financial stability for individuals with mental illnesses. Specifically, there are open research opportunities to develop technological interventions that leverage real-time, objective data to reduce risky financial decision-making for both individuals and institutions. We then describe existing technical, legal, and ethical challenges as well as providing a broad research agenda to address these challenges. Lastly, we will discuss how to promote equity and prevent discrimination for FinTech focusing on mental health support.

\subsection{Early-Warning Prediction}
Considering the complex relationship between mental health and finances, FinTech systems can help users learn about both their financial behaviors and their mental health status throughout time. Over the course of use, such systems could help predict when different mood states are about to happen and serve as an early warning for upcoming spending sprees. Conversely, the financial behaviors detected by a FinTech system could be used to predict the onset of future mood episodes. This would not only help users manage their mental health and financial decisions, but also explain the connections between the two \cite{giudici2021shapley}.

It is reasonable to believe that a system capable of early detection of symptomatic behaviors could have provided actionable feedback to our coauthor at the time of their episode, potentially affording improved self-awareness of risky behaviors and of illness states. This is especially salient considering the length of this episode and subtlety of its symptoms. Although the harms of this episode may be clear in hindsight, they remained opaque to the individual and several mental health professionals at the time. Objective financial data may have provided a strong signal of relapse.

However, we faced a number of challenges to present this case study in its present form. We believe many of these challenges represent requirements to operationalize objective financial data analysis in research settings. Central to these challenges is the theme of timely access to objective, structured financial data in machine-readable and human-readable formats. Considering our retrospective retrieval, our ability to obtain financial statements as PDF depended on the technical capacities and policies of the banking institution. If a longer period of time had elapsed beyond this symptomatic episode, the data retrieval steps might require processing paper-based records (i.e., digital data retrieval might be available only for a certain time period in some financial institutes). This would seriously hinder data collection at scale in retrospective studies.

Financial data collection methods will vary depending on the data source. At the scale of an N=1 case study, we could afford to manually process PDF statements. Larger studies may require utilizing more scalable data sources such as structured, delimited text or API-based access. Even at our scale, the use of PDF statements constrained our analysis steps. Many printed or digitized financial statements only list transactions by date. A finer-grained temporal analysis could yield valuable insights into behavioral changes during symptomatic states (e.g., how spending patterns change during disruptions to sleep). Additionally, our calculation of burstiness parameters was limited to a daily time-grain. More specific timestamps could yield useful, actionable insights into more specific patterns of spiky, rapid spending during symptomatic periods. Mood monitoring apps have also been found to relate to more formal measures of mood in bipolar disorder, suggesting that they might have utility in measuring mood \cite{chan2021jfr}. Linking such mobile based measures with live real-time measures of finances could therefore prove useful in the future.

Although reliance on retrospective patient self-report remains common in the field of psychiatry, our analysis was limited by a lack of specific labeling. Access to health records, granular self-reported mood logs, or sleep data would each individually improve our capacity to assess these financial data in more specific terms. In the following section, we will describe future research directions to identify early-warning signs by leveraging real-time, objective financial data.

\subsubsection{Research Directions}
Research in this domain remains limited by the scarcity of structured financial data to individuals and especially to trusted third-parties. The possibility of performing this research at scale depends on a standardized, permissioned technical infrastructure for electronic access by individuals and authorized third-parties. Although the technical infrastructure currently exists to support these aims, regulations are under development which may clarify the role of authorized third-party access to consumer financial data.

\paragraph{Better access to financial data}
As evidenced in our case study, objective financial data varies widely in quality and availability based on the data storage and access management practices of the individual’s banking institution. These variations pose a number of technical challenges, especially when considering how best to operationalize and scale data collection. Financial data can be made available in different formats including as paper or PDF documents, as structured text files, or using software-based approaches known as application programming interfaces, or APIs. We assess each of these types of source material by the quality and ease of access they each afford as shown in Table \ref{tab:factors}.

Document-based source material (printed or PDF financial statements) imply substantial preprocessing steps to render printed transaction-level detail as structured, digitized data. Transactions are likely logged on a daily basis. Metadata such as location or transaction category will likely be absent. Conditional inclusion or exclusion of transactions will likely only be possible after completion of preprocessing steps. Retrieval will likely involve manual processes and make real-time analysis impossible.

Files exported as delimited plain-text (e.g., from a web-based banking portal) are structured, but may require preprocessing steps to account for variations in institutional output. These may include more granular transaction-level data which is simpler to retrieve on a recurring basis. It may be possible to conditionally include or exclude specific transactions without revealing them to the researcher. Real-time analysis may be possible, but likely depends heavily on access protocols of the institution. Additional third-party metadata may be merged with these types of files.


Overall, API access (e.g., Plaid \cite{plaid}) supported by financial institutes provides real-time transaction-level financial data and allow for simple, permissioned, and revocable third-party access.

\begin{table}[]
\centering
\begin{tabular}{|l|l|l|l|l|l|l|}
\hline
Source & \textbf{\begin{tabular}[c]{@{}l@{}}Data \\ granularity\\  (e.g., \\ transaction \\ level, \\ day level)\end{tabular}} & \textbf{\begin{tabular}[c]{@{}l@{}}Processing\\ difficulty\\ (e.g., manual\\ to automatic)\end{tabular}} & \textbf{\begin{tabular}[c]{@{}l@{}}Metadata \\ available?\end{tabular}} & \textbf{Data structure} & \textbf{\begin{tabular}[c]{@{}l@{}}Possibility\\ of real-time\\ access?\end{tabular}} & \textbf{\begin{tabular}[c]{@{}l@{}}Retrieval \\ Difficulty\end{tabular}} \\ \hline
\textbf{\begin{tabular}[c]{@{}l@{}}PDF\\ Statement\end{tabular}} & \begin{tabular}[c]{@{}l@{}}Least \\ granular, \\ daily\\ time-grain\end{tabular} & \begin{tabular}[c]{@{}l@{}}Extremely \\ difficult to \\ scale\end{tabular} & \begin{tabular}[c]{@{}l@{}}No. Would\\ need \\ to manually \\ incorporate \\ third-party \\ data after \\ preprocessing\end{tabular} & \begin{tabular}[c]{@{}l@{}}Unstructured, \\ requires \\ preprocessing\end{tabular} & \begin{tabular}[c]{@{}l@{}}Non-realtime. \\ Impossible\\ to obtain \\ in realtime\end{tabular} & \begin{tabular}[c]{@{}l@{}}Manual \\ process, \\ difficult.\end{tabular} \\ \hline
\textbf{\begin{tabular}[c]{@{}l@{}}Retrospective \\ structured \\ data\end{tabular}} & \begin{tabular}[c]{@{}l@{}}Depends \\ on data \\ source\end{tabular} & \begin{tabular}[c]{@{}l@{}}Possible to \\ scale, \\ especially \\ retrospective\\ studies\end{tabular} & \begin{tabular}[c]{@{}l@{}}Possible, \\ but likely\\ requires \\ third-party\\ data\end{tabular} & \begin{tabular}[c]{@{}l@{}}Possible, \\ but likely \\ requires \\ preprocessing \\ steps\end{tabular} & \begin{tabular}[c]{@{}l@{}}Likely not \\ realtime, \\ but may be \\ possible\end{tabular} & \begin{tabular}[c]{@{}l@{}}Depends on \\ source \& \\ analysis \\ needs\end{tabular} \\ \hline
\textbf{\begin{tabular}[c]{@{}l@{}}API \\ integration\end{tabular}} & \begin{tabular}[c]{@{}l@{}}Per \\ transaction\end{tabular} & \begin{tabular}[c]{@{}l@{}}Immediately \\ scalable\end{tabular} & \begin{tabular}[c]{@{}l@{}}Possible, \\ given \\ institutional \\ capacity\end{tabular} & \begin{tabular}[c]{@{}l@{}}Structured, \\ machine- \\ and \\ human-\\ readable \\ data\end{tabular} & \begin{tabular}[c]{@{}l@{}}Possible, \\ given \\ institutional \\ support\end{tabular} & Simple \\ \hline
\end{tabular}
\caption{Factors to consider when collecting objective financial data.}
\label{tab:factors}
\end{table}

\paragraph{Integrating behavioral and contextual datastreams with financial data}
Future research should also explore how to integrate multimodal behavioral and contextual datastreams with financial data. This can lead to granular, context-specific understanding of how mental health symptoms can manifest in idiosyncratic financial behaviors.  For example, mood data collected using ecological momentary assessments (EMA) can be combined with financial transaction data. This can help to establish the relationship between emotional states and financial decision-making.

Financial data can also be combined with passively sensed data. Prior work has used behavioral and contextual data including location information to assess stability in bipolar disorder \cite{abdullah2016automatic,palmius2017itbe}. Combining such data streams with financial transaction information can provide valuable insights to researchers, clinicians, and individuals seeking tools for self-management and reflection. Furthermore, the availability of wearable devices and their ability to collect granular health data can complement financial data to predict early warning signs. For example, by leveraging data from wearable sleep-trackers, we can identify how circadian and sleep related factors might impact financial decision making and risk taking.

\subsection{Opportunities for Financial Data-driven  Interventions}
This use of objective financial data opens up new opportunities for financial intervention systems. In this section, we suggest future directions on different ways to intervene on point-of-purchase decisions and incorporate cognitive financial behavioral therapy activities to help improve overall financial wellbeing.

\subsubsection{Point of Purchase Interventions}
Online shopping has become streamlined and mostly frictionless. With auto-saved credit card information or one-step payment functions, as well as contactless purchasing, consumers spend less time on their purchasing decisions. As a result, steps in the purchasing process that would otherwise be an opportunity for the buyer to reconsider their decisions are often removed\cite{evans2017fintech}. While this lack of friction broadly impacts all users, this is particularly problematic for those prone to impulsive spending. Adding friction in online spending process can lead to harm reduction---both at an individual level and for banking institutions themselves. For instance, some UK banks have already added “positive friction” to help customers prevent gambling \cite{gamblingcomm}. Future interventions could help address impulsive spending and promote more mindful purchasing by adding increasing levels of friction with flexible settings. In other words, we can introduce friction into the process of purchasing based on the user’s changing situational needs or depending on their current mood state. 

To start, friction can be reintroduced into online purchasing, such as forcing the reentry of payment cards or adding two-factor authentication. While this ultimately does not prohibit the user from purchasing, it adds additional time to the activity which may slow down their decision process, make the user reconsider, and potentially choose not to follow through. In-situ reevaluation prompts could be provided to the user to help with informed decision making. Systems could ask the user questions such as “are you sure you want to buy X for \$Y?”to provide an opportunity to consider the potential impact of making this purchase. Constraints could also be added that items must sit in a virtual shopping cart for a certain amount of time (e.g., 24 or 48 hours) before the user can initiate payment---especially for very large purchases. Lastly, users could proactively set the system to restrict certain purchases when a mood episode is detected to prevent spending sprees. This could be tailored to the user by restricting purchases containing chosen keywords, by a specified cost threshold, or by time of day (e.g., limiting late-night impulse purchases). These restrictions could then be lifted under specific and predetermined conditions, such as detecting a mood episode has passed or having the user go through an attention or cognition-based assessment.

\subsubsection{Psychological Financial Therapies}
In collaboration with clinicians, we can also explore how to leverage existing psychological therapies such as Cognitive Behavioral Therapy (CBT) and Cognitive-Behavioral Financial Therapy (CBFT)\cite{nabeshima2015cognitive} to target impulsive spending behaviors and deliver them digitally. CBT has been delivered online and has been shown to be effective for depression and anxiety \cite {richards2015efficacy, wright2019computer}. CBT often involves individuals identifying and challenging unhelpful thinking patterns, as well as facing fears they are avoiding. With CBFT, this could be linked to financial difficulties, such as understanding the thinking patterns leading to impulse spending and trying to face finance-related fears. When incorporated with FinTech systems, users may be able to reflect on their behavior over time, recognize important patterns, and develop coping skills to manage their spending. Users could be prompted to evaluate and reflect on their behaviors during a set window and consider the context in which they made those financial decisions. Mindfulness has also been shown to predict later compulsive spending in bipolar disorder \cite{ richardson2019financial}, so incorporating this and helping individuals be more mindfully aware of the thoughts and feelings which might come before impulsive spending could be useful.

Future research should focus on developing just-in-time interventions to support in-situ and real-time decision making. For instance, if the system senses an increase in spending, the user could be pushed a suggestion to move funds to a savings account or make them temporarily inaccessible to them to preemptively reduce potential harm. Users could also be provided alternative activities, such as calling a friend or going for a walk, to prevent problematic behaviors in situations where spending is motivated by boredom or a want to elevate their current mood. In this case, these alternative activities could provide an outlet for their current needs and motivations behind spending that does not involve spending money. In a recent study, Richardson et al. \cite{richardson2022acceptability} developed `Space from Money Worries' --- an online CBT-based intervention designed to tackle the link between financial difficulties and mental health problems, including impulsive spending. Their data showed reduced symptoms of depression and anxiety, as well as lowered financial distress \cite{richardson2022acceptability}. Linking such interventions with objective financial data could help determine whether they have a real-world impact on spending patterns and other financial behaviors.

\subsection{Challenges for FinTech to Support Mental Health}
While objective financial data opens up these opportunities to support financial wellbeing and intervene on problematic behaviors, there are a number of challenges to address through future research, including user agency, privacy, and equity concerns.

\subsubsection{Personal Autonomy and User Agency}

The use of financial data for mood state prediction and intervention presents several open research questions for personal autonomy and user agency. Given that sensitive data can be used to limit or control users’ behavior, it is important to ask who has access to and ownership of the data, as well as the power to make decisions for the user---whether that is the system itself or their care partners. These systems can also empower users to take back control of their financial wellbeing in the long run. However, given the nature of impulsive spending and different mood states, there are times when user agency has to be limited or handed-off to other to support their overall goals of harm-reduction.

Simply put, some users might not want to step back from the ledge on their own when faced with impulsive spending opportunities, especially when in specific mood states. Some users may have an awareness of such behavioral patterns about themselves, but others may not. As previous work has shown, those who are long aware of this, often take it upon themselves to proactively curb high volume spending \cite{richardson2019financial}. For these users, maintaining control of their own financial management systems may be sufficient, as they can set up appropriate spending limitation settings on their own when they anticipate needing them. Other high-risk users may be less aware of their mood states and their relationship to spending sprees. These users may benefit from giving up some of their agency in return for more support. This could take the form of the system acting on their behalf in certain circumstances, like if it senses mood episode onset, or transferring some of the decision-making power to other people.

\paragraph{Research directions}
Moving forward, we should work to understand personal risk thresholds to better determine what type of intervention is most suitable for individual users. This personal risk threshold could be used to determine whether other people should be involved in users’ financial decisions, and what level of friction is the best fit to slow down their impulsive spending habits. An ideal system should strive for a balance between autonomy, constraint and individual risk profiles. To work out this ideal balance, future work should be oriented around the following questions: What level of friction is appropriate and effective? For whom? And who decides this? What level of autonomy is needed and when? Our overall goal should be to establish how we can help promote financial stability for vulnerable individuals while preserving their personal autonomy as much as possible.

\paragraph{Supporting Shared and Collaborative Decision Making}

Previous work has leveraged open banking data to notify pre-selected care partners of different financial events (e.g., low balance or high-dollar alerts) \cite{barrospenaPick2021} to improve financial wellbeing. For those with higher risk thresholds, one could argue that some users might benefit from more active third-party involvement, including the delegation of power to transact and the disclosure of financial information. In this case, care partners could play a more active role in decision making and co-managing patient finances. For instance, a payment system could be set to require care partner approval before significant purchases are authorized. 

A collaborative approach can help address some of overarching agency concerns. By using a more collaborative system, control can be distributed across a team of individuals when user agency needs to be limited. This would involve working within their larger social support network, rather than transferring financial decision-making power to one individual. In this case, primary users would work with their support networks---clinician, family, or care partners---to make a preemptive plan of action for specific situations. This could leverage existing clinical practices (e.g., \emph{advanced directives}) \cite{SAMHSA} to define the conditions in which the users agency in financial decisions are limited. Ideally, mental capacity would be assessed prior to big purchases or taking out loans, as when individuals are in a manic mood episodes they may legally lack the mental capacity to consent to those decisions. However, this approach could dictate what decisions can be made on their behalf, by whom, for how long, and under what circumstances this plan goes into effect. While this type of agreement is traditionally a legal document drafted between patients and clinicians, it could then be carried out and reinforced through FinTech system architecture.

\subsubsection{Privacy}
For both institutional and individual level interventions, it is crucial to consider privacy risks and concerns with using objective banking data for financial intervention and ensure the systems we develop remain in the users’ best interest \cite{morton2021using}. Privacy also needs to be considered in how the outputs from this system are presented, used, and potentially shared with others. The main goal here should be to find the right balance between providing the most useful, accurate information to benefit the user while also maintaining as much of their privacy as possible. In some circumstances, users may want to share their data with other members of their social support networks. For this, it could be beneficial to grant information access to others at different levels of granularity, based on their roles in the larger system. For example, a therapist could have access to long-term behavioral data for clinical purposes. Whereas, a friend or family member may only be provided “need to know '' information or a task by task basis, should the user need assistance with different financial decisions.

Given the amount of social stigma that surrounds both mental health conditions and poor finances, the data used to support these systems is highly sensitive and personal to users. If users are uncomfortable with the amount of data used, how their privacy is protected, and the amount of trust given to the system, they are unlikely to use the system long enough to see any personal benefits. To support user adoption and long term use, it is highly important that we ensure privacy to maintain user trust in the intervention system. Future work is needed to understand user acceptance and the privacy trade-offs involved with using personal data within this context. 

\paragraph{Establish privacy and acceptance concerns}
Through qualitative work we can gather a more in-depth understanding of the range of privacy concerns different users may have with using objective, real-time data to monitor and intervene on their spending behaviors. This work could be carried out through interviews with individuals with mental health issues, their family members and care partners, and their clinical support. By collecting data from multiple stakeholders, we can gain in depth insights into concerns and expectations when it comes to developing FinTech to support mental health issues. Furthermore, providing a prototype for users to explore during these sessions can also prompt a more thorough discussion on privacy, trust, and shared decision making. This could give users a more concrete idea of the kind of data involved, but also what users stand to get out of the system. 

\paragraph{Establish data sharing policies}
Additional work could look at the wider context in which these systems might be used, for the purpose of establishing ideal data sharing policies. Focus groups with primary users and members of their social support networks could be conducted to gather a better understanding of these concerns and expectations of all parties involved. Similarly, scenario-based data collection could gather insight on the specific situations where users would be comfortable sharing their data with others and at what granularity. These next steps would provide a better sense of privacy tradeoffs that users may work through and help design future FinTech systems in alignment with their specific privacy needs.

\paragraph{Privacy preserving analysis}
Recent technical developments may allow for financial mental health research to be conducted while preserving the privacy of participants. Federated learning, for example, creates the possibility of ``training statistical models over remote devices or siloed data centers, such as mobile phones or hospitals, while keeping data localized" \cite{li2020ispm}. In this case, a user's financial data could be securely stored and modeled on their own smartphone without ever being exposed to third-parties. Further, users who wish to grant third parties access to their financial data using API-based methods have a high level of control over what is shared. Third party access could be restricted to certain accounts or types of transactions, to transactions within certain date ranges, or based on certain financial conditions being met. An individual's trusted care partner could have a different level of access than their accountant, for example. Perhaps most importantly, an individual sharing their financial data using API-based methods also has the option to revoke this access.

\subsubsection{Equity and Preventing Discrimination}
FinTech systems can operative on either an institutional or individual level, each providing users differential access to resources. Therefore, equity and discrimination should also be considered when developing interventions. We should prioritize making these systems more accessible to primary users---those who need them for help with their own finances, by promoting digital access and increased financial citizenship and limiting possible problematic use.

\paragraph{Institutional vs. Individual approaches} 
With an institutional approach, financial health interventions can intervene at the point of individual transactions and control overall money outflow from the user’s bank account. These efforts could reduce harm for the user, such as decreasing their likelihood of defaulting on loans or going into bankruptcy. This could also prevent banks from granting loans to those who are not in a state, financially and psychologically, to take on these high-risk financial decisions. This helps the individual avoid negative financial consequences of defaulting on these loans in the future, which is also a significant incentive for the bank itself \cite{giudici2018fintech}, who would otherwise not see repayment.

Conversely, individual-level FinTech systems could provide assistance outside the confines of existing banking institutions and potentially help a wider range of users. Rather than relying on financial transaction data from banking institutions, an individual-level system could instead use online purchasing data or other alternative sources. An individual-level system could then provide financial cognitive behavioral therapies \cite{nabeshima2015cognitive} reinforced through system architecture, involve existing social support networks to help the user with financial decisions and meet behavior change goals---outside the confines of banking institutions.

\paragraph{Equity and Discrimination}
Previous work in this space has largely focused on institutional approaches. The lack of individual-level options limits who can benefit from them. While institutional level interventions involving banks can help promote good money habits, this is only accessible to those with existing financial citizenship, who already have the financial means to maintain a traditional bank account. Focusing solely on institutional-level interventions helps those with privilege of existing banking access avoid future bankruptcies, foreclosures, defaulting on loans, or other dire financial situations---potentially leaving out those in the most need \cite{farr2019financial}. 

Individual-level interventions can also present similar concerns when data or control is shared or handed off to others. Choosing these third-party allies may be a difficult decision for users to make. Some may feel obligated to choose those who are closest to them, like a parent or a partner, who may expect to be given this role, even if the user does not feel comfortable with it. Others may lack significant relationships or experience strained relationships, leaving them with no or few options for trusted support. These scenarios may lead users to hand over control to those who may not use it in their best interest. Additional work is needed to uncover the range of challenges that may present in this context, as well as how to better support users in the process of choosing and maintaining their financial allies.

Future work is needed to address institutional and individual equity concerns. At an institutional level, it is important to consider the bank’s role in the overall system. This could mean limiting how much of this personal data is accessible to banks to only the minimum amount necessary for the intervention to function as designed. Shifting too much control to banks could give them the power to assume all Fintech users lack the ability to make their own financial decisions, regardless of individual circumstances or current mood state, and deny opportunities unjustly. More research is needed to understand the ideal role for banking institutions within the wider system and how to balance the needs of both the institution and individual users. Future work should also expand on individual-level options to provide more equitable systems that operate independently from banking institutions. Improving access to financial data for individuals would take us one step closer to this, as open financial data is currently limited, especially in the United States.

\section{Conclusion}
Mental illness and financial instability are often linked with each other. Financial woes can be detrimental to mental health and problematic decision making during symptomatic periods can lead to long-term financial ruin. Recent studies have explored the bidirectional relationship between mental illnesses and financial instability. However, they have been mostly limited to self-reported data, which might not provide a complete picture due to recall bias and coarse sampling methods. This lack of granular understanding is also a serious barrier toward developing effective, just-in-time interventions to support financial stability for individuals with mental illnesses. In recent years, there has been an increasing focus on open and accessible financial data (e.g., Open Banking API). Several third-party tools and platforms now provide access to transaction and payment data across different financial institutes. The resultant granular and (near) real-time access to financial data can help to identify nuanced relationships between financial (in)stability and mental health issues as well as developing personalized, just-in-time interventions. However, there are considerable challenges before we can use objective financial data for mental health support.

In this paper, we aim to explore opportunities in identifying financial ``biomarkers” and intervention strategies for individuals with mental illnesses and document their associated challenges for future research. We have identified potentially risky financial behavioral dynamics across symptomatic phases (e.g., ratio of increased credit spending). While we have not found statistically significant trends nor our findings are generalizable beyond this case, our approach provides an insight into the challenges of accessing objective financial data. Based on our findings, we have also explored potential intervention strategies. Furthermore, we have described technical, ethical, and equity challenges in developing financial data-driven assessment and intervention methods, as well as providing a broad research agenda to address these challenges. We believe that being able to leverage objective, personalized financial data in a privacy-preserving and ethical manner will lead to a paradigm shift in mental health care.

\section*{Conflict of Interest Statement}
T. R. has been paid as a consultant and receives royalties from a software company for developing an intervention around financial difficulties and mental health. TR also acts as an advisor to a financial technology company for which he receives shares.


\bibliographystyle{frontiersinHLTH&FPHY} 
\bibliography{bibliography}

\begin{thebibliography}{44}
\expandafter\ifx\csname natexlab\endcsname\relax\def\natexlab#1{#1}\fi
\expandafter\ifx\csname urlstyle\endcsname\relax
  \expandafter\ifx\csname doi\endcsname\relax
  \def\doi#1{doi:\discretionary{}{}{}#1}\fi \else
  \expandafter\ifx\csname doi\endcsname\relax
  \def\doi{doi:\discretionary{}{}{}\begingroup \urlstyle{rm}\Url}\fi \fi
\expandafter\ifx\csname selectlanguage\endcsname\relax
  \def\selectlanguage#1{}\fi

\bibitem[{Trautmann et~al.(2016)Trautmann, Rehm, and
  Wittchen}]{trautmann2016economic}
Trautmann S, Rehm J, Wittchen HU.
\newblock The economic costs of mental disorders: Do our societies react
  appropriately to the burden of mental disorders?
\newblock {\em EMBO reports\/} {\bf 17} (2016) 1245--1249.

\bibitem[{Richardson et~al.(2013)Richardson, Elliott, and
  Roberts}]{richardson2013relationship}
Richardson T, Elliott P, Roberts R.
\newblock The relationship between personal unsecured debt and mental and
  physical health: a systematic review and meta-analysis.
\newblock {\em Clinical psychology review\/} {\bf 33} (2013) 1148--1162.

\bibitem[{Saraceno and Barbui(1997)}]{saraceno1997poverty}
Saraceno B, Barbui C.
\newblock Poverty and mental illness.
\newblock {\em The Canadian Journal of Psychiatry\/} {\bf 42} (1997) 285--290.

\bibitem[{APA(2013)}]{american2013dsm}
APA.
\newblock Dsm-5 diagnostic classification.
\newblock {\em Diagnostic and statistical manual of mental disorders\/} {\bf
  10} (2013).

\bibitem[{Jones et~al.(2015)Jones, Metcalf, Gordon-Smith, Forty, Perry, Lloyd
  et~al.}]{jones2015gambling}
Jones L, Metcalf A, Gordon-Smith K, Forty L, Perry A, Lloyd J, et~al.
\newblock Gambling problems in bipolar disorder in the uk: prevalence and
  distribution.
\newblock {\em The British Journal of Psychiatry\/} {\bf 207} (2015) 328--333.

\bibitem[{Fletcher et~al.(2013)Fletcher, Parker, Paterson, and
  Synnott}]{fletcher2013high}
Fletcher K, Parker G, Paterson A, Synnott H.
\newblock High-risk behaviour in hypomanic states.
\newblock {\em Journal of affective disorders\/} {\bf 150} (2013) 50--56.

\bibitem[{Richardson et~al.(2017{\natexlab{a}})Richardson, Jansen, Turton, and
  Bell}]{richardson2017relationship}
Richardson T, Jansen M, Turton W, Bell L.
\newblock The relationship between bipolar disorder and financial difficulties:
  A qualitative exploration of client’s views.
\newblock {\em Clinical Psychology Forum\/} (2017{\natexlab{a}}).

\bibitem[{Richardson et~al.(2018)Richardson, Jansen, and
  Fitch}]{richardson2018financial}
Richardson T, Jansen M, Fitch C.
\newblock Financial difficulties in bipolar disorder part 1: Longitudinal
  relationships with mental health.
\newblock {\em Journal of Mental Health\/} {\bf 27} (2018) 595--601.

\bibitem[{Richardson et~al.(2015)Richardson, Elliott, Waller, and
  Bell}]{richardson2015longitudinal}
Richardson T, Elliott P, Waller G, Bell L.
\newblock Longitudinal relationships between financial difficulties and eating
  attitudes in undergraduate students.
\newblock {\em International Journal of Eating Disorders\/} {\bf 48} (2015)
  517--521.

\bibitem[{Richardson et~al.(2017{\natexlab{b}})Richardson, Elliott, Roberts,
  and Jansen}]{richardson2017longitudinal}
Richardson T, Elliott P, Roberts R, Jansen M.
\newblock A longitudinal study of financial difficulties and mental health in a
  national sample of british undergraduate students.
\newblock {\em Community mental health journal\/} {\bf 53} (2017{\natexlab{b}})
  344--352.

\bibitem[{Inc.(2021)}]{plaid}
Inc P.
\newblock Plaid: Enabling all companies to build fintech solutions  (2021).
\newblock Available at \url{https://plaid.com/}.

\bibitem[{Morton et~al.(2021)Morton, Torous, Murray, and
  Michalak}]{morton2021using}
Morton E, Torous J, Murray G, Michalak EE.
\newblock Using apps for bipolar disorder--an online survey of healthcare
  provider perspectives and practices.
\newblock {\em Journal of Psychiatric Research\/} {\bf 137} (2021) 22--28.

\bibitem[{Marazziti et~al.(2010)Marazziti, Consoli, Picchetti, Carlini, and
  Faravelli}]{marazziti2010cognitive}
Marazziti D, Consoli G, Picchetti M, Carlini M, Faravelli L.
\newblock Cognitive impairment in major depression.
\newblock {\em European journal of pharmacology\/} {\bf 626} (2010) 83--86.

\bibitem[{Cheema et~al.(2015)Cheema, MacQueen, and
  Hassel}]{cheema2015assessing}
Cheema MK, MacQueen GM, Hassel S.
\newblock Assessing personal financial management in patients with bipolar
  disorder and its relation to impulsivity and response inhibition.
\newblock {\em Cognitive neuropsychiatry\/} {\bf 20} (2015) 424--437.

\bibitem[{Richardson et~al.(2019)Richardson, Jansen, and
  Fitch}]{richardson2019financial}
Richardson T, Jansen M, Fitch C.
\newblock Financial difficulties in bipolar disorder part 2: psychological
  correlates and a proposed psychological model.
\newblock {\em Journal of Mental Health\/}  (2019).

\bibitem[{Alexander et~al.(2017)Alexander, Oliver, Burdine, Tang, and
  Dunlop}]{alexander2017reported}
Alexander LF, Oliver A, Burdine LK, Tang Y, Dunlop BW.
\newblock Reported maladaptive decision-making in unipolar and bipolar
  depression and its change with treatment.
\newblock {\em Psychiatry research\/} {\bf 257} (2017) 386--392.

\bibitem[{Park and Sela(2018)}]{park2018not}
Park JJ, Sela A.
\newblock Not my type: Why affective decision makers are reluctant to make
  financial decisions.
\newblock {\em Journal of Consumer Research\/} {\bf 45} (2018) 298--319.

\bibitem[{Narayan et~al.(2011)Narayan, Case, and Edwards}]{narayan2011role}
Narayan B, Case DO, Edwards SL.
\newblock The role of information avoidance in everyday-life information
  behaviors.
\newblock {\em Proceedings of the American Society for Information Science and
  Technology\/} {\bf 48} (2011) 1--9.

\bibitem[{Granero et~al.(2016)Granero, Fern{\'a}ndez-Aranda, Mestre-Bach,
  Steward, Ba{\~n}o, del Pino-Guti{\'e}rrez et~al.}]{granero2016compulsive}
Granero R, Fern{\'a}ndez-Aranda F, Mestre-Bach G, Steward T, Ba{\~n}o M, del
  Pino-Guti{\'e}rrez A, et~al.
\newblock Compulsive buying behavior: clinical comparison with other behavioral
  addictions.
\newblock {\em Frontiers in Psychology\/} {\bf 7} (2016) 914.

\bibitem[{Ram{\'\i}rez-Mart{\'\i}n et~al.(2020)Ram{\'\i}rez-Mart{\'\i}n,
  Ramos-Mart{\'\i}n, Mayoral-Cleries, Moreno-K{\"u}stner, and
  Guzman-Parra}]{ramirez2020impulsivity}
Ram{\'\i}rez-Mart{\'\i}n A, Ramos-Mart{\'\i}n J, Mayoral-Cleries F,
  Moreno-K{\"u}stner B, Guzman-Parra J.
\newblock Impulsivity, decision-making and risk-taking behaviour in bipolar
  disorder: a systematic review and meta-analysis.
\newblock {\em Psychological medicine\/}  (2020) 1--13.

\bibitem[{Cook et~al.(2004)Cook, Burke, Petersen, Carter, Reinheimer, Cohran
  et~al.}]{cook2004assessing}
Cook JA, Burke JK, Petersen CA, Carter T, Reinheimer C, Cohran L, et~al.
\newblock Assessing the financial planning needs of americans with mental
  illnesses.
\newblock {\em University of Illinois at Chicago Center on Mental Health
  Services Research and Policy\/}  (2004).

\bibitem[{Tharp et~al.(2016)Tharp, Johnson, Sinclair, and
  Kumar}]{tharp2016goals}
Tharp JA, Johnson SL, Sinclair S, Kumar S.
\newblock Goals in bipolar i disorder: Big dreams predict more mania.
\newblock {\em Motivation and Emotion\/} {\bf 40} (2016) 290--299.

\bibitem[{Tovanich et~al.(2021)Tovanich, Centellegher, Bennacer~Seghouani,
  Gladstone, Matz, and Lepri}]{tovanich2021eds}
Tovanich N, Centellegher S, Bennacer~Seghouani N, Gladstone J, Matz S, Lepri B.
\newblock Inferring psychological traits from spending categories and dynamic
  consumption patterns.
\newblock {\em EPJ Data Science\/} {\bf 10} (2021) 1--23.
\newblock \doi{10.1140/epjds/s13688-021-00281-y}.

\bibitem[{Leverich and Post(2002)}]{leverich2002}
Leverich G, Post R.
\newblock The {NIMH} {Life} {Chart} {Manual}. for {Recurrent} {Affective}
  {Illness}: {The} {LCM} - {S}/{R} {Retrospective} ({Self}-{Version})  (2002).

\bibitem[{Honig et~al.(2001)Honig, Hendriks, Akkerhuis, and
  Nolen}]{honig2001peac}
Honig A, Hendriks CH, Akkerhuis GW, Nolen WA.
\newblock Usefulness of the retrospective {Life}-{Chart} method manual in
  outpatients with a mood disorder: a feasibility study.
\newblock {\em Patient Education and Counseling\/} {\bf 43} (2001) 43--48.
\newblock \doi{10.1016/S0738-3991(00)00144-0}.

\bibitem[{cam(2021)}]{camelot}
{Camelot}: {PDF} {Table} {Extraction} for {Humans}  (2021).

\bibitem[{pan(2021)}]{pandas}
pandas - {Python} {Data} {Analysis} {Library}  (2021).

\bibitem[{Liu et~al.(2008)Liu, Ting, and Zhou}]{liu2008}
Liu FT, Ting KM, Zhou ZH.
\newblock Isolation {{Forest}}.
\newblock {\em 2008 {{Eighth IEEE International Conference}} on {{Data
  Mining}}\/} (2008), 413--422.
\newblock \doi{10.1109/ICDM.2008.17}.

\bibitem[{Pedregosa et~al.(2011)Pedregosa, Varoquaux, Gramfort, Michel,
  Thirion, Grisel et~al.}]{scikit-learn}
Pedregosa F, Varoquaux G, Gramfort A, Michel V, Thirion B, Grisel O, et~al.
\newblock Scikit-learn: Machine learning in {P}ython.
\newblock {\em Journal of Machine Learning Research\/} {\bf 12} (2011)
  2825--2830.

\bibitem[{Giudici and Raffinetti(2021)}]{giudici2021shapley}
Giudici P, Raffinetti E.
\newblock Shapley-lorenz explainable artificial intelligence.
\newblock {\em Expert Systems with Applications\/} {\bf 167} (2021) 114104.

\bibitem[{Chan et~al.(2021)Chan, Sun, Aitchison, and Sivapalan}]{chan2021jfr}
Chan EC, Sun Y, Aitchison KJ, Sivapalan S.
\newblock Mobile {App}–{Based} {Self}-{Report} {Questionnaires} for the
  {Assessment} and {Monitoring} of {Bipolar} {Disorder}: {Systematic} {Review}.
\newblock {\em JMIR Formative Research\/} {\bf 5} (2021) e13770.
\newblock \doi{10.2196/13770}.

\bibitem[{Abdullah et~al.(2016)Abdullah, Matthews, Frank, Doherty, Gay, and
  Choudhury}]{abdullah2016automatic}
Abdullah S, Matthews M, Frank E, Doherty G, Gay G, Choudhury T.
\newblock Automatic detection of social rhythms in bipolar disorder.
\newblock {\em Journal of the American Medical Informatics Association\/} {\bf
  23} (2016) 538--543.

\bibitem[{Palmius et~al.(2017)Palmius, Tsanas, Saunders, Bilderbeck, Geddes,
  Goodwin et~al.}]{palmius2017itbe}
Palmius N, Tsanas A, Saunders KEA, Bilderbeck AC, Geddes JR, Goodwin GM, et~al.
\newblock Detecting {Bipolar} {Depression} {From} {Geographic} {Location}
  {Data}.
\newblock {\em IEEE Transactions on Biomedical Engineering\/} {\bf 64} (2017)
  1761--1771.
\newblock \doi{10.1109/TBME.2016.2611862}.
\newblock Conference Name: IEEE Transactions on Biomedical Engineering.

\bibitem[{Evans and Action(2017)}]{evans2017fintech}
Evans K, Action R.
\newblock Fintech for good; how financial technology can support people
  experiencing mental health problems.
\newblock {\em London: The Money and Mental Health Policy Institute. Available
  at: http://www. moneyandmentalhealth.
  org/wpcontent/uploads/2017/07/Fintech-for-good-report. pdf\/}  (2017).

\bibitem[{gam(2021)}]{gamblingcomm}
Block gambling payments with your bank.
\newblock {\em UK Gambling Commission\/}  (2021).

\bibitem[{Nabeshima and Klontz(2015)}]{nabeshima2015cognitive}
Nabeshima G, Klontz BT.
\newblock Cognitive-behavioral financial therapy.
\newblock {\em Financial Therapy\/} (Springer) (2015), 143--159.

\bibitem[{Richards et~al.(2015)Richards, Richardson, Timulak, and
  McElvaney}]{richards2015efficacy}
Richards D, Richardson T, Timulak L, McElvaney J.
\newblock The efficacy of internet-delivered treatment for generalized anxiety
  disorder: A systematic review and meta-analysis.
\newblock {\em Internet Interventions\/} {\bf 2} (2015) 272--282.

\bibitem[{Wright et~al.(2019)Wright, Owen, Richards, Eells, Richardson, Brown
  et~al.}]{wright2019computer}
Wright JH, Owen JJ, Richards D, Eells TD, Richardson T, Brown GK, et~al.
\newblock Computer-assisted cognitive-behavior therapy for depression: a
  systematic review and meta-analysis.
\newblock {\em The Journal of clinical psychiatry\/} {\bf 80} (2019) 0--0.

\bibitem[{Richardson et~al.(2022)Richardson, Enrique, Earley, Adegoke, Hiscock,
  and Richards}]{richardson2022acceptability}
Richardson T, Enrique A, Earley C, Adegoke A, Hiscock D, Richards D.
\newblock The acceptability and initial effectiveness of “space from money
  worries”: An online cognitive behavioral therapy intervention to tackle the
  link between financial difficulties and poor mental health.
\newblock {\em Frontiers in Public Health\/}  (2022) 716.

\bibitem[{Barros~Pena et~al.(2021)Barros~Pena, Kursar, Clarke, Alpin, Holkar,
  and Vines}]{barrospenaPick2021}
Barros~Pena B, Kursar B, Clarke RE, Alpin K, Holkar M, Vines J.
\newblock ``{{Pick Someone Who Can Kick Your Ass}}" - {{Moneywork}} in
  {{Financial Third Party Access}}.
\newblock {\em Proceedings of the ACM on Human-Computer Interaction\/} {\bf 4}
  (2021) 218:1--218:28.
\newblock \doi{10.1145/3432917}.

\bibitem[{SAMHSA(2019)}]{SAMHSA}
SAMHSA.
\newblock A practical guide to psychiatric advance directives  (2019).

\bibitem[{Li et~al.(2020)Li, Sahu, Talwalkar, and Smith}]{li2020ispm}
Li T, Sahu AK, Talwalkar A, Smith V.
\newblock Federated {Learning}: {Challenges}, {Methods}, and {Future}
  {Directions}.
\newblock {\em IEEE Signal Processing Magazine\/} {\bf 37} (2020) 50--60.
\newblock \doi{10.1109/MSP.2020.2975749}.
\newblock Conference Name: IEEE Signal Processing Magazine.

\bibitem[{Giudici(2018)}]{giudici2018fintech}
Giudici P.
\newblock Fintech risk management: A research challenge for artificial
  intelligence in finance.
\newblock {\em Frontiers in Artificial Intelligence\/}  (2018) 1.

\bibitem[{Farr et~al.(2019)Farr, Cash, and Harper}]{farr2019financial}
Farr B, Cash B, Harper A.
\newblock Why financial institutions need to offer supportive banking features
  (2019).

\end{thebibliography}

\end{document}